\documentclass[11pt,twocolumn]{article}
\usepackage{graphicx}
\usepackage{amsmath}
\usepackage{bm}

\newcommand\T{\rule{0pt}{2.6ex}}

\begin{document}

\title{\textsf{\textbf{A geometric algebra reformulation of 2x2 matrices:\\ the dihedral group $\mathcal D_4$ in bra-ket notation}}}
\author{Quirino M. Sugon Jr,* Carlo B. Fernandez, and Daniel J. McNamara
\smallskip\\
\small{Ateneo de Manila University, Department of Physics, Loyola Heights, Quezon City, Philippines 1108}\\
\small{*Also at Manila Observatory, Upper Atmosphere Division, Ateneo de Manila University Campus}\\
\small{e-mail: \texttt{qsugon$@$observatory.ph}}}
\date{\small{\today}}
\maketitle

\small
\section*{}\label{Abstract}
\textbf{Abstract.}  
We represent vector rotation operators in terms of bras or kets of half-angle exponentials in Clifford (geometric) algebra $\mathcal Cl_{3,0}$.  We show that $SO_3$ is a rotation group and we define the dihedral group $D_4$ as its finite subgroup. We use the Euler-Rodrigues formulas to compute the multiplication table of $\mathcal D_4$ and derive its group algebra identities.  We take the linear combination of rotation operators in $\mathcal D_4$ to represent the four Fermion matrices in Sakurai, which in turn we use to decompose any $2\times 2$ matrix.  We show that bra and ket operators generate left- and right-acting matrices, respectively.  We also show that the Pauli spin matrices are not vectors but vector rotation operators, except for $\hat\sigma_2$ which requires a subsequent multiplication by the imaginary number $i$ geometrically interpreted as the unit oriented volume.

\section{Introduction}

Clifford (geometric) algebra $\mathcal Cl_{3,0}$ may be presented by replacing the three Cartesian basis vectors and the unit real number by their correponding $2\times2$ Pauli spin matrices\cite{Hestenes_1971_ajp39i9pp1013-1027_p1018}.  For example,
\begin{eqnarray}
\label{eq:unit is sigma_0}
1&\equiv&\hat\sigma_0=
\begin{pmatrix}
1&0\\
0&1
\end{pmatrix}
,\\
\label{eq:e_1 is sigma_1}
\mathbf e_1&\equiv&\hat\sigma_1=
\begin{pmatrix}
0&1\\
1&0
\end{pmatrix}
,\\
\label{eq:e_2 is sigma_2}
\mathbf e_2&\equiv&\hat\sigma_2=
\begin{pmatrix}
0&-i\\
i&0
\end{pmatrix}
,\\
\label{eq:e_3 is sigma_3}
\mathbf e_3&\equiv&\hat\sigma_3=
\begin{pmatrix}
1&0\\
0&-1
\end{pmatrix}
.
\end{eqnarray}

From these definitions we can show that the orthonormality relation holds\cite{BaylisHuschiltWei_1992_ajp60i9pp788-797_p789}:
\begin{equation}
\label{eq:orthonormality intro}
\mathbf e_j\mathbf e_k+\mathbf e_k\mathbf e_j=2\delta_{jk},
\end{equation}
for $j,k\in\{1,2,3\}$.  And that the product of two vectors $\mathbf a$ and $\mathbf b$ is governed by the Pauli identity\cite{Hestenes_2003_ajp71i2pp104-121_p110}
\begin{equation}
\label{eq:Pauli identity intro}
\mathbf a\mathbf b=\mathbf a\cdot\mathbf b+i(\mathbf a\times\mathbf b),
\end{equation}
where $i=\mathbf e_1\mathbf e_2\mathbf e_3$ is an imaginary number that commutes with vectors.

The representation of the unit vectors in terms of matrices presupposes that matrices are more fundamental than vectors.  But what if we adopt the opposite view that vectors are more fundamental than matrices?  That is, given only the rule for multiplying unit vectors in Eq.~(\ref{eq:orthonormality intro}), can we arrive at a definition of a matrix and its corresponding algebra?

Yes, we can.  To justify this claim requires three steps.  First, we represent the symmetry group of a square, the dihedral group $\mathcal D_4$, by exponentials of imaginary half vectors as done in the following general form for vector rotation\cite{Vold_1993_ajp61i6pp491-504_p499}:
\begin{equation}
\label{eq:general vector rotation intro}
\mathbf r'=e^{-i\bm\theta/2}\,\mathbf r\,e^{i\bm\theta/2},
\end{equation}
where $\bm\theta$ is the axis of rotation of vector $\mathbf r$ and $|\bm\theta|$ is the magnitude of counterclockwise angle of rotation.  Second, we construct linear combinations of the $\mathcal D_4$ rotation operators to represent the four Fermion matrices in Sakurai\cite{Sakurai_1967_AdvancedQuantumMechanics_p80} or the four standard basis for dioptric power space in Harris\cite{Harris_1997_OptomVisSci76i6pp349-366_p357}:
\begin{eqnarray}
\label{eq:e_11 and e_12 matrix intro}
\textsf e_{11}=
\begin{pmatrix}
1&0\\
0&0
\end{pmatrix}
,&\quad&
\textsf e_{12}=
\begin{pmatrix}
0&1\\
0&0
\end{pmatrix}
,
\\
\label{eq:e_21 and e_22 matrix intro}
\textsf e_{21}=
\begin{pmatrix}
0&0\\
1&0
\end{pmatrix}
,&\quad&
\textsf e_{22}=
\begin{pmatrix}
0&0\\
0&1
\end{pmatrix}
.
\end{eqnarray}
And third, we express any matrix as a linear combination of the four Fermion matrices.

\begin{figure}[hb]
\begin{center}
\setlength{\unitlength}{1 mm}
\begin{picture}(40,60)(-20,-30)
\thicklines
\qbezier(-20,0)(20,0)(20,0)
\thinlines
\qbezier(0,0)(15,20)(15,20)
\qbezier(15,18)(15,20)(15,20)
\qbezier(13,19)(15,20)(15,20)
\put(17,21){$\mathbf r$}
\qbezier(-15,-20)(-13.800,-18.400)(-13.800,-18.400)
\qbezier(-12.600,-16.800)(-11.400,-15.200)(-11.400,-15.200)
\qbezier(-10.200,-13.600)(-9.000,-12.000)(-9.000,-12.000)
\qbezier(-7.800,-10.400)(-6.600,-8.800)(-6.600,-8.800)
\qbezier(-5.400,-7.200)(-4.200,-5.600)(-4.200,-5.600)
\qbezier(-3.000,-4.000)(0,0)(0,0)
\qbezier(-2,-1)(0,0)(0,0)
\qbezier(0,-2)(0,0)(0,0)
\put(-19,-24){$\mathbf r$}
\qbezier(0,0)(15,-20)(15,-20)
\qbezier(15,-18)(15,-20)(15,-20)
\qbezier(13,-19)(15,-20)(15,-20)
\put(15,-24){$-\bm\eta\mathbf r\bm\eta$}
\qbezier(0,0)(-15,20)(-15,20)
\qbezier(-15,18)(-15,20)(-15,20)
\qbezier(-13,19)(-15,20)(-15,20)
\put(-23,22){$\bm\eta\mathbf r\bm\eta$}
\qbezier(0,0)(0,20)(0,20)
\qbezier(1,18)(0,20)(0,20)
\qbezier(-1,18)(0,20)(0,20)
\put(-1,22){$\bm\eta$}
\end{picture}
\end{center}
\begin{quote}
\vspace{-0.5cm}
\caption{\footnotesize The vector $\bm\eta\mathbf r\bm\eta$ is the ray $\mathbf r$ flipped $180^\circ$ about the mirror normal $\bm\eta$.  The opposite of $\bm\eta\mathbf r\bm\eta$ is the reflected ray $-\bm\eta\mathbf r\bm\eta$.}
\label{fig:flips and reflections}
\vspace{-0.5cm}
\end{quote}
\end{figure}
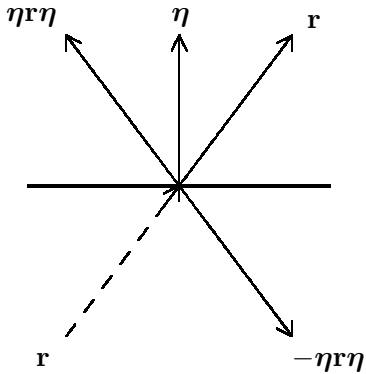

In Hestenes's representation of symmetry groups, the fundamental operation is not rotation but reflection.  If $\mathbf r$ is a vector reflected with respect to the mirror with normal vector unit vector $\bm\eta$, then\cite{Hestenes_2002_AGACSEpp3-34_p7}
\begin{equation}
\label{eq:r' is -nrn intro}
\mathbf r'=\underline N\mathbf r=-\bm\eta\mathbf r\bm\eta,
\end{equation}
where $\underline N$ is a left-acting reflection operator.  The product of two reflections is a rotation\cite{Hestenes_2002_AGACSEpp3-34_p8}:
\begin{equation}
\label{eq:reflection N_1 N_2 r intro}
\underline N_1\underline N_2\mathbf r=(\bm\eta_2\bm\eta_1)\,\mathbf r\,(\bm\eta_1\bm\eta_2)=e^{-i\bm\theta/2}\mathbf \,\mathbf r\,e^{i\bm\theta/2},
\end{equation}
where $\bm\theta$ is parallel to $\bm\eta_1\times\bm\eta_2$ and $|\bm\theta|/2$ is the angle between $\bm\eta_1$ and $\bm\eta_2$.  (See Fig.~\ref{fig:flips and reflections})

In our case, we shall view rotations as more fundamental than reflections.  To do this, we need to define a new operation similar to but distinct from reflection: flip.  If $\mathbf r'$ is the vector $\mathbf r$ flipped with respect to the axis along the unit vector $\bm\eta$, then
\begin{equation}
\label{eq:r' is r ket in intro}
\mathbf r'=\bm\eta\mathbf r\bm\eta,
\end{equation}    
which differs from that of reflection by a sign.  Notice that a flip is actually a $180^\circ$ rotation about $\bm\eta$:
\begin{equation}
\label{eq:r' is r ket in pi over 2 intro}
\mathbf r'=e^{-i\bm\eta\pi/2}\,\mathbf r\,e^{i\bm\eta\pi/2}=(-i\bm\eta)\mathbf r(i\bm\eta)=\bm\eta\mathbf r\bm\eta.
\end{equation}

Hestenes remarked that the pair of rotors (rotation operators) $\pm R$ distinguish equivalent rotations in opposite senses\cite{Hestenes_2002_AGACSEpp3-34_p7}.  In Hestenes underbar notation, we may write this as
\begin{equation}
\label{eq:x' is x R is R dagger x R pm intro}
\mathbf x'=\mathbf x\underline{\pm R}=(\pm R)^\dagger\mathbf x(\pm R),
\end{equation}
where $R=e^{i\bm\theta/2}$ ($\underline{\pm R}$ is actually left-acting in Hestenes, but we changed its direction of action for consistency with our conventions).  This theorem of Hestenes is very useful, so we shall rederive it here.  His underline notation, however, we shall change to bras or kets, in order to suggest the direction of action:
\begin{equation}
\label{eq:x' is x R is bra R x is x ket R}
\mathbf x'=\mathbf x\underline R=\langle R^\dagger|\mathbf x=\mathbf x|R\rangle.
\end{equation}
Notice that the bars suggest some similarity with the \emph{absolute value} bars $|\ |$, which is what we intend.  Note that the $\langle R^\dagger|$ and $|R\rangle$ are the bra and ket representations of the rotation operator; these representations are only notations for they do not obey Dirac's bracket algebra as described in Simmons and Guttmann\cite{SimmonsGuttmann_1970_StatesWavesandPhotons_p48-49}.  

We shall divide this paper into five sections.  The first section is the Introduction.  In the second section, we shall give a brief summary of Clifford (geometric) algebra $\mathcal Cl_{3,0}$.  We shall discuss the orthonormality axiom, the Pauli identity for vector products, the Euler's theorem for the exponential of an imaginary vector, and the general vector rotation formula.  In the third section, we shall use the exponential rotation operators in geometric algebra to describe the elements of the general rotation group in three dimensions, $SO_3$.  We shall rederive Rodrigues formulas for the composition of rotations to prove the group properties of $SO_3$.  We shall also rederive Hestenes theorem on equivalent rotations, together with some theorems on bra and ket transformation.  In the the fourth section, we discuss the rotation subgroup $\mathcal D_4$, which is a finite subgroup of $SO_3$.  We shall construct the $\mathcal D_4$ group algebra and use this to define the Fermion and Pauli $2\times 2$ matrices.  The fifth section is Conclusions.

\section{Geometric Algebra}

\subsection{Vectors and Imaginary Numbers}

The Clifford (geometric) algebra $\mathcal Cl_{3,0}$ is an associative algebra generated by three vectors ${\mathbf e}_1$, ${\mathbf e}_2$, and ${\mathbf e}_3$ that satisfy the orthonormality relation in Eq.~(\ref{eq:orthonormality intro}).  That is, 
\begin{eqnarray}
\label{eq:e_j squared is 1}
\mathbf e_j^2 &=& 1,\\
\label{eq:e_je_k is -e_ke_j}
\mathbf e_j\mathbf e_k&=&-\mathbf e_k\mathbf e_j,\quad j\neq k.
\end{eqnarray}
We shall refer to Eqs.~(\ref{eq:e_j squared is 1}) and (\ref{eq:e_je_k is -e_ke_j}) as the normality and orthogonality axioms, respectively.

Let ${\mathbf a}$ and ${\mathbf b}$ be two vectors spanned by the three unit spatial vectors in $\mathcal Cl_{3,0}$.  By the orthonormality axioms in Eqs. (\ref{eq:e_j squared is 1}) and (\ref{eq:e_je_k is -e_ke_j}), we can show that the product of these two vectors is given by the Pauli identity in Eq.~(\ref{eq:Pauli identity intro}):
\begin{equation}
\label{eq:Pauli identity}
\mathbf a\mathbf b=\mathbf a\cdot\mathbf b+i(\mathbf a\times\mathbf b).
\end{equation}
  Notice that if $\mathbf a_\parallel$ and $\mathbf a_\perp$ are the components of $\mathbf a$ parallel and perpendicular to $\mathbf b$, then 
\begin{eqnarray}
\label{eq:a parallel b is b a parallel}
\mathbf a\cdot\mathbf b&=&\mathbf a_\parallel\mathbf b=\mathbf b\,\mathbf a_\parallel,\\
\label{eq:a perp b is - b a perp}
i(\mathbf a\times\mathbf b)&=&\mathbf a_\perp\mathbf b=\mathbf b\,\mathbf a_\perp.
\end{eqnarray}
That is, parallel vectors commute; perpendicular vectors anticommute.

In general, we may express every element $\hat A$ in $\mathcal Cl_{3,0}$ as a linear combination of a scalar, a vector, an imaginary vector (bivector), and an imaginary scalar (trivector):
\begin{equation}
\label{eq:cliffor A}
\hat A = A_0 + {\mathbf A}_1 +i{\mathbf A}_2 + iA_3.
\end{equation}
Following Jancewicz, we shall call $\hat A$ as a cliffor\cite{Jancewicz_1988_MultivectorstoCliffordAlgebrainElectrodynamics_p28}.

\subsection{Rotations}

Let $\bm\theta$ be a vector.  Multiplying this by $i$ results to the bivector $i\bm\theta$, which we may geometrically interpret as the oriented plane perpendicular to vector $\bm\theta$ (or in the language of forms, $\bm\theta$ is the Hodge map or dual of $i\bm\theta$).  It is easy to see that the square of $i\bm\theta$ is negative:
\begin{equation}
\label{eq:i theta squared}
(i\bm\theta)^2=i^2{\bm\theta}^2=-|{\bm\theta}|^2.
\end{equation}
Thus, we may use Euler's theorem in complex analysis to write
\begin{equation}
\label{eq:Euler theorem}
e^{i\bm\theta}=\cos|{\bm\theta}|+i\frac{\bm\theta}{|{\bm\theta}|}\sin|{\bm\theta}|.
\end{equation}
Notice that the exponential $e^{i\bm\theta}$, like the product $\mathbf a\mathbf b$ in Eq.~(\ref{eq:Pauli identity}), is a sum of a scalar and an imaginary vector.

\begin{figure}[h]
\begin{center}
\setlength{\unitlength}{1 mm}
\begin{picture}(60,80)(-28,-45)
\qbezier(30,0)(34.330,7.500)(21.495,11.250)
\qbezier(21.495,11.250)(8.660,15.000)(-8.505,11.250)
\qbezier(-8.505,11.250)(-25.670,7.500)(-30.000,0.000)
\thicklines
\qbezier(-30.000,0.000)(-34.330,-7.500)(-21.495,-11.250)
\qbezier(-21.495,-11.250)(-8.660,-15.000)(8.505,-11.250)
\qbezier(8.505,-11.250)(25.670,-7.500)(30.000,0.000)
\thinlines
\qbezier(0.000,-12.603)(0,-40)(0,-40)
\qbezier(0,0)(0,-4)(0,-4)
\qbezier(0,-6)(0,-8)(0,-8)
\qbezier(0,-10)(0,-12)(0,-12)
\qbezier(0,0)(1,-2)(1,-2)
\qbezier(0,0)(-1,-2)(-1,-2)
\put(-5,-23){$\mathbf r_\parallel$}
\qbezier(3.632,-1.670)(5.454,-0.946)(6.000,0.000)
\put(10,-3.7){$\theta$}
\qbezier(0,0)(0,25)(0,25)
\qbezier(-1,23)(-1,23)(0,25)
\qbezier(1,23)(1,23)(0,25)
\put(-1,27){$\bm\theta$}
\qbezier(0,0)(18.160,-8.350)(18.160,-8.350)
\qbezier(16,-8.5)(18.160,-8.350)(18.160,-8.350)
\qbezier(18,-7)(18.160,-8.350)(18.160,-8.350)
\put(7,-8){$\mathbf r_\perp$}
\qbezier(0,0)(30,0)(30,0)
\qbezier(28,1)(30,0)(30,0)
\qbezier(28,-1)(30,0)(30,0)
\put(15,2){$\mathbf r\,'_\perp$}
\qbezier(0,-40)(18.160,-8.350)(18.160,-8.350)
\qbezier(18,-11)(18.160,-8.350)(18.160,-8.350)
\put(8,-22){$\mathbf r$}
\qbezier(0,-40)(30,0)(30,0)
\qbezier(30,-2)(30,0)(30,0)
\put(20,-18){$\mathbf r\,'$}
\end{picture}
\end{center}
\begin{quote}
\vspace{-0.5cm}
\caption{\footnotesize The vector $\mathbf r'=e^{-i\bm\theta/2}\,\mathbf r\ e^{i\bm\theta/2}$ is the vector $\mathbf r$ rotated counterclockwise about the vector $\bm\theta$ by an angle $\theta=|\bm\theta|$.}
\label{fig:general vector rotation}
\vspace{-1cm}
\end{quote}
\end{figure}
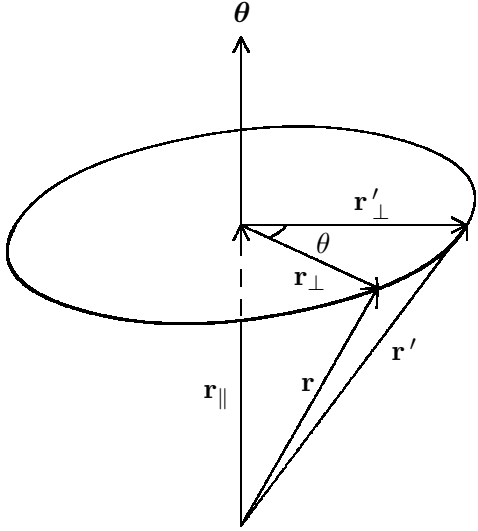

If $\mathbf r'$ is the vector $\mathbf r$ rotated counterclockwise about the vector ${\bm\theta}$ by an angle $|{\bm\theta}|$ (Figure~\ref{fig:general vector rotation}), we write
\begin{equation}
\label{eq:r prime is exp -itheta2 r exp itheta2}
\mathbf r=e^{-i{\bm\theta}/2}\;\mathbf r\;e^{i{\bm\theta}/2}.
\end{equation}
If we expand $\mathbf r$ in terms of its components parallel and perpendicular to $\bm\theta$, then we can show that
\begin{eqnarray}
\label{eq:r parallel exp i theta commute}
\mathbf r_\parallel e^{i{\bm\theta}/2}&=&e^{i{\bm\theta}/2}\mathbf r_\parallel,\\
\label{eq:r perp exp i theta anticommute}
\mathbf r_\perp e^{i{\bm\theta}/2}&=&e^{-i{\bm\theta}/2}\mathbf r_\perp,
\end{eqnarray}
so that Eq.~(\ref{eq:r prime is exp -itheta2 r exp itheta2}) reduces to
\begin{equation}
\label{eq:r prime is r parallel + r perp exp i theta}
\mathbf r^\prime=\mathbf r_\parallel+\mathbf r_\perp e^{i\bm\theta}.
\end{equation}
Thus, the parallel component is unaffected by the rotation.  (See Fig.~\ref{fig:general vector rotation})

The second term on the right side of Eq.~(\ref{eq:r prime is r parallel + r perp exp i theta}) may be expanded as
\begin{eqnarray}
\label{eq:r exp i theta is r cos theta + i r theta sin theta}
\mathbf r_\perp e^{i\bm\theta}&=&\mathbf r_\perp\cos|{\bm\theta}|+i\mathbf r_\perp\frac{\bm\theta}{|\bm\theta|}\sin|\bm\theta|\nonumber\\
&=&\mathbf r_\perp\cos|{\bm\theta}|-\mathbf r_\perp\times\frac{\bm\theta}{|\bm\theta|}\sin|\bm\theta|,
\end{eqnarray}
where we used the Pauli identity in Eq.~(\ref{eq:Pauli identity}).  Equation~(\ref{eq:r exp i theta is r cos theta + i r theta sin theta}) states that $\mathbf r_\perp e^{i\bm\theta}$ is the vector $\mathbf r_\perp$ rotated counterclockwise about the vector $\bm\theta$ by an angle $|\bm\theta|$.\cite{SugonMcNamara_2004_ajp72i1pp92-97_p93}

Rotation operators may also act on a cliffor $\hat A$ to yield a new cliffor $\hat A'$:
\begin{equation}
\label{eq:A' is rotate A}
\hat A'=e^{-i\bm\theta/2}\hat A\,e^{i\bm\theta/2}
\end{equation}
Using the expansion of $\hat A$ in Eq.~(\ref{eq:cliffor A}), together with the commutation relations
\begin{eqnarray}
\label{eq:A_0 exp commute}
A_0\, e^{i\bm\theta/2}=e^{i\bm\theta/2}A_0,\\
\label{eq:iA_3 exp commute}
iA_3\, e^{i\bm\theta/2}=e^{i\bm\theta/2}iA_3,
\end{eqnarray}
we arrive at
\begin{eqnarray}
\label{eq:A' is rotate A scalar-trivector}
A_0'+iA_3'&=&A_0+iA_3,\\
\label{eq:A' is rotate A vector-bivector}
\mathbf A_1'+i\mathbf A_2'&=&e^{-i\bm\theta/2}(\mathbf A_1+i\mathbf A_2)e^{i\bm\theta/2},
\end{eqnarray}
after separating the linearly independent parts.  Thus only the vector-bivector part is affected by the rotation; the scalar-trivector (complex number) remains unchanged.

\section{The Rotation Group $SO_3$}

\subsection{Group Properties}

Rotations form a group labelled as $SO(3)$ or the special orthogonal group in three dimensions (orthogonal matrices are traditionally used to represent the group elements).  To prove that $SO_3$ is a group, we must show that it satisfies four properties: closure, associativity, existence of an inverse, and existence of an identity.\cite{Hestenes_1990_NewFoundationsforClassicalMechanics_p298}  
\medskip

\textbf{a.  Closure.}  The product of two rotations is also a rotation.  That is, if $\bm\theta_1$ and $\bm\theta_2$ are vectors, there exists a vector $\bm\theta_3$ such that
\begin{equation}
\label{eq:Rodrigues theorem}
e^{-i\bm\theta_3/2}\,\mathbf r\,e^{i\bm\theta_3/2}=e^{-i\bm\theta_2/2}e^{-i\bm\theta_1/2}\,\mathbf r\,e^{i\bm\theta_1/2}e^{i\bm\theta_2/2},
\end{equation}
where $\mathbf r$ is a vector in $\mathcal Cl_{3,0}$. Equation~(\ref{eq:Rodrigues theorem}) is known as Rodrigues's theorem. 

To prove the claim in Eq.~(\ref{eq:Rodrigues theorem}), it is sufficient to prove only the equality of the right exponentials,
\begin{equation}
\label{eq:exp theta_3 is exp theta_1 exp theta_2}
e^{i\bm\theta_3/2}=e^{i\bm\theta_1/2}e^{i\bm\theta_2/2},
\end{equation}
because the equality of the left exponentials immediately follows by taking the inverse of Eq.~(\ref{eq:exp theta_3 is exp theta_1 exp theta_2}):
\begin{equation}
\label{eq:exp -theta_3 is exp -theta_2 exp -theta_1}
e^{-i\bm\theta_3/2}=e^{-i\bm\theta_2/2}e^{-i\bm\theta_1/2}.
\end{equation}

Expanding the exponentials in Eq.~(\ref{eq:exp theta_3 is exp theta_1 exp theta_2}), distributing the terms, and separating the scalar and imaginary vector parts, we obtain 
\begin{eqnarray}
\label{eq:Rodrigues formula cos}
\cos\frac{|\bm\theta_3|}{2}&=&\cos\frac{|\bm\theta_1|}{2}\cos\frac{|\bm\theta_2|}{2}\nonumber\\
& &-\ \frac{\bm\theta_1\cdot\bm\theta_2} {|\bm\theta_1||\bm\theta_2|}\sin\frac{|\bm\theta_1|}{2}\sin\frac{|\bm\theta_2|}{2},
\end{eqnarray}
\begin{eqnarray}
\label{eq:Rodrigues formula sin}
\frac{\bm\theta_3}{|\bm\theta_3|}\sin\frac{|\bm\theta_3|}{2}&=&\frac{\bm\theta_1}{|\bm\theta_1|}\sin\frac{|\bm\theta_1|}{2}\cos\frac{|\bm\theta_2|}{2}\nonumber\\
& &+\ \frac{\bm\theta_2}{|\bm\theta_2|}\sin\frac{|\bm\theta_2|}{2}\cos\frac{|\bm\theta_1|}{2}\nonumber\\
& &-\ \frac{\bm\theta_1\times\bm\theta_2}{|\bm\theta_1||\bm\theta_2|}\sin\frac{|\bm\theta_2|}{2}\sin\frac{|\bm\theta_1|}{2},
\end{eqnarray}
where we factored out $i$ in the second equation.  We shall refer to these equations as the Euler-Rodrigues's formulas.\cite{Hestenes_1990_NewFoundationsforClassicalMechanics_pp282-284}\cite{Hladik_1999_SpinorsinPhysics_p15}

We can solve for the magnitude $|\bm\theta_3|$ by taking the inverse cosine of Eq.~(\ref{eq:Rodrigues formula cos}).  The value of $|\bm\theta_3|$ is unique in the range $0\leq|\bm\theta_3|\leq\pi$.  After solving for $|\bm\theta_3|$, we then substitute this result in Eq.~(\ref{eq:Rodrigues formula sin}) to solve for the unique direction angle $\bm\theta_3/|\bm\theta_3|$.  Thus, given vectors $\bm\theta_1$ and $\bm\theta_2$, there exists a vector $\bm\theta_3$ defined by Eqs.~(\ref{eq:Rodrigues formula cos}) and (\ref{eq:Rodrigues formula sin}), so that Eq.~(\ref{eq:exp theta_3 is exp theta_1 exp theta_2}) holds, which is what we wish to show.  

\textbf{b.  Associativity.} Because products in $\mathcal Cl_{3,0}$ are associative, then 
\begin{eqnarray}
\label{eq:rotations associativity property}
\mathbf r'\!\!&=&\!\!\!\!(e^{-i\bm\theta_3/2}e^{-i\bm\theta_2/2})e^{-i\bm\theta_1/2}\,\mathbf r\,e^{i\bm\theta_1/2}(e^{i\bm\theta_2/2}e^{i\bm\theta_3/2}),\nonumber\\
&=&\!\!\!\!e^{-i\bm\theta_3/2}(e^{-i\bm\theta_2/2}e^{-i\bm\theta_1/2})\,\mathbf r\,(e^{i\bm\theta_1/2}e^{i\bm\theta_2/2})e^{i\bm\theta_3/2}.
\end{eqnarray}
Thus, products of rotations are associative.

\textbf{c.  Identity.}  The identity rotation exists:
\begin{equation}
\label{eq:rotations identity property}
\mathbf r=e^{-0}\,\mathbf r\,e^0.
\end{equation}

\textbf{d.  Inverse.}  Every rotation has an inverse.  That is, if $\bm\theta$ is a vector, then
\begin{eqnarray}
\label{eq:rotations inverse property}
\mathbf r&=&e^{i\bm\theta/2}e^{-i\bm\theta/2}\,\mathbf r\,e^{i\bm\theta/2}e^{-i\bm\theta/2}\nonumber\\
&=&e^{-i\bm\theta/2}e^{i\bm\theta/2}\,\mathbf r\,e^{-i\bm\theta/2}e^{i\bm\theta/2}.
\end{eqnarray}

\subsection{Bra and Ket Representations}

Let us rewrite the rotation expression in Eq.~(\ref{eq:r prime is exp -itheta2 r exp itheta2}) as
\begin{equation}
\label{eq:r prime is exp -itheta2 r exp itheta2 bra ket}
\mathbf r=e^{-i{\bm\theta}/2}\;\mathbf r\;e^{i{\bm\theta}/2}=\langle e^{-i{\bm\theta}/2}|\mathbf r=\mathbf r|e^{i{\bm\theta}/2}\rangle,
\end{equation}
where
\begin{eqnarray}
\label{eq:bra is exp half theta}
\langle\,e^{-i\bm\theta/2}\,|=\left\langle\cos\frac{\bm\theta}{2}-i\frac{\bm\theta}{|\bm\theta|}\sin\frac{|\bm\theta|}{2}\,\right|,\\
\label{eq:ket is exp half theta}
|\,e^{i\bm\theta/2}\,\rangle=\left|\ \cos\frac{\bm\theta}{2}+i\frac{\bm\theta}{|\bm\theta|}\sin\frac{|\bm\theta|}{2}\,\right\rangle.
\end{eqnarray}
We shall refer to the angle braces expressions as the bra and ket representations for an element of a rotation group, respectively.  The representations on the left side of the equations are the polar form; those on the right side are the Cartesian form.  Note that the bra is a left-acting operator, while the ket is a right-acting one.

The use of bra and ket representations can be tricky.  So let us summarize some useful identities and comment on some likely algebraic traps:
\medskip

\textbf{a.  Action on Cliffors.}  
The bra and ket operators can also act on a general cliffor $\hat A$:
\begin{equation}
\label{eq:bra A is A ket}
\langle\, e^{-i\bm\theta/2}\,|\,\hat A=\hat A\,|\,e^{i\bm\theta/2}\,\rangle.
\end{equation}
If $\hat A=A_0+iA_3$, then $\hat A$ is unaffected by the rotation:
\begin{equation}
\label{eq:bra A_0 + iA_3 is A_0 + iA_3 ket}
A_0+iA_3=\langle e^{-i\bm\theta/2}|(A_0+iA_3)=(A_0+iA_3)|\,e^{i\bm\theta/2}\,\rangle.
\end{equation}

Now, since the ket $|e^{i\bm\theta/2}\rangle$ is a right-acting operator and the bra $\langle e^{-i\bm\theta/2}|$ is a left-acting one, we must not put the cliffor operand $\hat A$ on the right side of a ket operator or on the left side of a bra operator.  But if we do so on purpose, the result is a new operator:
\begin{eqnarray}
\label{eq:r ket A is exponential expand}
\mathbf r\,|\,e^{i\bm\theta/2}\,\rangle\hat A=e^{-i\bm\theta/2}\,\mathbf r\,e^{i\bm\theta/2}\hat A,\\
\label{eq:A bra r is exponential expand}
\hat A\langle\, e^{-i\bm\theta/2}\,|\,\mathbf r=\hat A\, e^{-i\bm\theta/2}\,\mathbf r\, e^{i\bm\theta/2}.
\end{eqnarray}
That is, the action of the new operator is as follows: rotate the vector $\mathbf r$ about the vector $\bm\theta$ counterclockwise by an angle $|\bm\theta|$ and the result multiply by $\hat A$.

What if we have a chain of cliffors and kets:
\begin{equation}
\label{eq:r A ket chain}
\mathbf r\,\hat A_1\,|K_1\rangle\hat A_2\,|K_2\rangle\cdots\hat A_n\,|K_n\rangle?
\end{equation}
How shall we interpret this?  Let us adopt the following convention: if $\hat A\,|K\rangle$ is \emph{explicitly} defined as a right-acting operator acting on $\mathbf r$, then $\hat A$ acts first on $\mathbf r$ and the result is acted on by $|K\rangle$:
\begin{equation}
\label{eq:r A ket order}
\mathbf r\,\hat A\,|K\rangle =(\mathbf r\hat A)|K \rangle.
\end{equation}
We shall call this the left-most precedence convention for right-acting operators (for left-acting operators it is right-most precedence).  Hence,
\begin{eqnarray}
\label{eq:r A ket chain order}
& &\mathbf r\,\hat A_1\,|K_1\rangle\hat A_2\,|K_2\rangle\cdots\hat A_n\,|K_n\rangle\nonumber\\
& &\qquad\qquad= (\cdots(((\mathbf r\hat A_1)|K_1\rangle)\hat A_2)\cdots\hat A_n)|K\rangle.
\end{eqnarray}

\textbf{b.  Equivalent Kets.} We may rewrite each counterclockwise rotation operator in terms of its clockwise counterpart:  
\begin{equation}
\label{eq:rotation counterclockwise clockwise}
|\,e^{i\bm\theta/2}\,\rangle=|\,e^{-i\mathbf e_{\theta} (2\pi-|\bm\theta|)/2}\,\rangle,
\end{equation}
where
\begin{equation}
\label{eq:e_theta}
\mathbf e_\theta=\frac{\bm\theta}{|\bm\theta|},
\end{equation}
which should not be confused with the unit vector in polar coordinates.  Because
\begin{equation}
\label{eq:exp i e_theta pi is -1}
e^{\pm i\mathbf e_\theta\pi}=-1,
\end{equation}
then Eq.~(\ref{eq:rotation counterclockwise clockwise}) reduces to
\begin{equation}
\label{eq:rotation counterclockwise clockwise reduced}
|\,e^{i\bm\theta/2}\,\rangle=|\,-e^{i\bm\theta/2}\,\rangle.
\end{equation}
Thus, two representations are equivalent if their corresponding ket arguments are additive inverses of each other. (We may think of the ket bars as absolute value operators.)  We shall refer to Eq.~(\ref{eq:rotation counterclockwise clockwise reduced}) as the Hestenes's rotation representation equivalence.

\textbf{c.  Products of Kets.}  The product of two kets is the ket of the product of their arguments:
\begin{equation}
\label{eq:product of kets}
|\,e^{i\bm\theta_1/2}\,\rangle |\,e^{i\bm\theta_2/2}\,\rangle=|\,e^{i\bm\theta_1/2}e^{i\bm\theta_1/2}\,\rangle,
\end{equation}
because
\begin{equation}
\label{eq:product of kets exponential explanation}
\mathbf r\, |\,e^{i\bm\theta_1/2}\,\rangle |\,e^{i\bm\theta_2/2}\,\rangle=e^{-i\bm\theta_2/2}e^{-i\bm\theta_1/2}\,\mathbf r\,e^{i\bm\theta_1/2}e^{i\bm\theta_2/2}.
\end{equation}
For example, the factorization
\begin{equation}
\label{eq:i e_2 exp i e_1 pi over 4 factorization}
|\, e^{i\mathbf e_2\pi/2} e^{i\mathbf e_1\pi/4}\,\rangle
=|\,i \mathbf e_2\,\rangle |\,e^{i\mathbf e_1\pi/4}\,\rangle
\end{equation}
is allowed, because $i\mathbf e_2$ is expressible as an exponential of a imaginary vector.  On the other hand, 
\begin{equation}
\label{eq:ie_2 exp ie_1 pi over 4 factorization false}
|\, e^{i\mathbf e_2\pi/2} e^{i\mathbf e_1\pi/4} \,\rangle \neq |\, i \,\rangle |\, \mathbf e_2e^{i\mathbf e_1\pi/4} \,\rangle,
\end{equation}
because $i$ and ${\mathbf e_2}e^{i\mathbf e_1\pi/4}$ cannot be similarly expressed.  

To avoid confusion, we shall not write bra and ket operators in the same side of an equation as in the expectation value expression in Quantum Mechanics,
\begin{eqnarray}
\label{eq:r' is bra r ket}
\mathbf r'=\langle e^{-i\bm\theta_1/2}|\,\mathbf r\,|e^{i\bm\theta_2/2}\rangle&=\!\!\!\!\!^?&\ \mathbf r\,|e^{i\bm\theta_1/2}\rangle|e^{i\bm\theta_2/2}\rangle\nonumber\\
&=\!\!\!\!\!^?&\ \mathbf r\,|e^{i\bm\theta_2/2}\rangle|e^{i\bm\theta_1/2}\rangle,
\end{eqnarray}
for we would not know which operator comes first.  As a convention, we shall use ket operators in our computations and convert our final results in terms of bra operators, but only if necessary.

\textbf{d.  Sum of Kets.} The rule for the sum of two kets is simple: do not combine or split their arguments.  That is,
\begin{equation}
\label{eq:sum of representations}
|\,e^{i\bm\theta_1}\,\rangle+|\,e^{i\bm\theta_2}\,\rangle\neq|\,e^{i\bm\theta_2}+e^{i\bm\theta_2}\,\rangle,
\end{equation}
unless, of course, if one of the terms is zero.  The reason for this rule is in the original definition of rotations in Eq.~(\ref{eq:Rodrigues theorem}), where we derived our half-angle representation for right acting operators.  That is, Eq.~(\ref{eq:sum of representations}) is not allowed because
\begin{eqnarray}
e^{-i\bm\theta_1/2}\!\!\!\!&\mathbf r&\!\!\!\!e^{i\bm\theta_1/2}+e^{-i\bm\theta_2/2}\ \mathbf r\ e^{i\bm\theta_2/2}\nonumber\\
&\neq&\!\!(e^{-i\bm\theta_2/2}+e^{-i\bm\theta_1/2})\,\mathbf r\,(e^{i\bm\theta_1/2}+e^{i\bm\theta_2/2}).
\end{eqnarray}

\section{Rotation Group $\mathcal D_4$}
\subsection{Group Elements}

\begin{figure}[ht]
\begin{center}
\setlength{\unitlength}{1 mm}
\begin{picture}(70,75)(-30,-32.5)
\multiput(21.213,-21.213)(-42.426,0){2}{\line(0,1){42.426}}
\multiput(-21.213,21.213)(0,-42.426){2}{\line(1,0){42.426}}
\thicklines
\qbezier(0,0)(30,0)(30,0)
\qbezier(28,1)(30,0)(30,0)
\qbezier(28,-1)(30,0)(30,0)
\put(32,-1){$\mathbf e_1$}
\qbezier(0,0)(0,30)(0,30)
\qbezier(-1,28)(0,30)(0,30)
\qbezier(1,28)(0,30)(0,30)
\put(-1,32){$\mathbf e_2$}
\qbezier(0,0)(21.213,21.213)(21.213,21.213)
\qbezier(20.213,18.213)(21.213,21.213)(21.213,21.213)
\qbezier(18.213,20.213)(21.213,21.213)(21.213,21.213)
\put(23,22){$\mathbf e_{1+2}$}
\qbezier(0,0)(21.213,-21.213)(21.213,-21.213)
\qbezier(20.213,-18.213)(21.213,-21.213)(21.213,-21.213)
\qbezier(18.213,-20.213)(21.213,-21.213)(21.213,-21.213)
\put(23,-24){$\mathbf e_{1-2}$}
\end{picture}
\end{center}
\begin{quote}
\vspace{-0.5cm}
\caption{\footnotesize The symmetry axes of a unit square in the $xy-$plane.  The vector $\mathbf e_3$ is pointing out of the paper.}
\label{fig:square symmetry axes}
\vspace{-0.5cm}
\end{quote}
\end{figure}
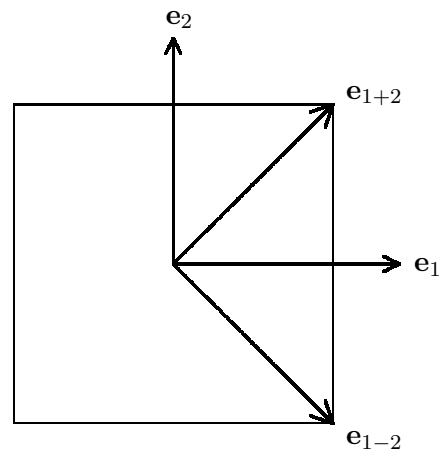

Let us define the positions of the vertices of a square to be the set 
\begin{equation}
\label{eq:vertices V_4 of a square}
\mathcal V_4=\{\mathbf e_{1+2},\mathbf e_{-1+2},\mathbf e_{-1+2},\mathbf e_{1-2}\},
\end{equation}
where
\begin{equation}
\label{eq:e_pm 1 pm 2}
\mathbf e_{\pm 1\pm 2}=\frac{1}{\sqrt 2}(\pm\mathbf e_1\pm\mathbf e_2).
\end{equation}
Notice that the symmetry axes of the square are
\begin{equation}
\label{eq:S_4 symmetry axes}
\mathcal S_4=\{\mathbf e_1,\mathbf e_2,\mathbf e_3,\mathbf e_{1+2},\mathbf e_{1-2}\}.
\end{equation}
(See Figure~\ref{fig:square symmetry axes})

If we define the dihedral group $\mathcal D_4$ to be the set of symmetry operations on a square defined in Eq.~(\ref{eq:vertices V_4 of a square}), then we may represent the elements of $\mathcal D_4$ as
\begin{eqnarray}
\label{eq:D_4 identity 1}
1&=&|\,1\,\rangle,\\
\label{eq:D_4 e_1 pi}
|\,e^{i\mathbf e_1\pi/2}\,\rangle&=&|\,i\mathbf e_1\,\rangle=|\,e^{-i\mathbf e_1\pi/2}\,\rangle,\\
\label{eq:D_4 e_2 pi}
|\,e^{i\mathbf e_2\pi/2}\,\rangle&=&|\,i\mathbf e_2\,\rangle=|\,e^{-i\mathbf e_2\pi/2}\,\rangle,\\
\label{eq:D_4 e_3 pi}
|\,e^{i\mathbf e_3\pi/2}\,\rangle&=&|\,i\mathbf e_3\,\rangle=|\,e^{-i\mathbf e_3\pi/2}\,\rangle,\\
\label{eq:D_4 e_3 pi half}
|\,e^{i\mathbf e_3\pi/4}\,\rangle&=&|\,\mathbf e_{0+3i}\,\rangle=|\,e^{-i\mathbf e_33\pi/4}\,\rangle,\\
\label{eq:D_4 e_3 -pi half}
|\,e^{-i\mathbf e_3\pi/4}\,\rangle&=&|\,\mathbf e_{0-3i}\,\rangle=|\,e^{i\mathbf e_33\pi/4}\,\rangle,\\
\label{eq:D_4 e_1 + e_2}
|\,e^{i(\mathbf e_{1+2})\pi/2}\,\rangle&=&|\,i\mathbf e_{1+2}\,\rangle=|\,e^{-i(\mathbf e_{1+2})\pi/2}\,\rangle,\\
\label{eq:D_4 e_1 - e_2}
|\,e^{i(\mathbf e_{1\!-2})\pi/2}\,\rangle&=&|\,i\mathbf e_{1\!-2}\,\rangle=|\,e^{-i(\mathbf e_{1\!-2})\pi/2}\,\rangle,
\end{eqnarray}
where
\begin{equation}
\label{eq:e_0 pm 3i}
\mathbf e_{0\pm 3i}=\frac{1}{\sqrt 2}(1\pm i\mathbf e_3).
\end{equation}
Note that we have removed the $\pm$ sign in the arguments of the representations, because the two representations are equivalent by Hestenes's theorem in Eq.~(\ref{eq:rotation counterclockwise clockwise reduced}).  For example, we could have written Eqs.~(\ref{eq:D_4 identity 1}) and (\ref{eq:D_4 e_1 pi}) as
\begin{eqnarray}
\label{eq:D_4 identity 1 pm}
1&=&|\pm 1\,\rangle,\\
\label{eq:D_4 e_1 pi pm}
|\pm e^{i\mathbf e_1\pi/2}\,\rangle&=&|\pm i\mathbf e_1\,\rangle=|\pm e^{-i\mathbf e_1\pi/2}\,\rangle.
\end{eqnarray}

There are only two generators for $\mathcal D_4$: 
\begin{equation}
G_{\mathcal D4}=\{\ |\,i\mathbf e_1\,\rangle,|\,\mathbf e_{0+3i}\,\rangle\}.
\end{equation}
The first corresponds to a flip about $\mathbf e_1$; the second, to a $\pi/2$ counterclockwise rotation about $\mathbf e_3$.  The other group elements can be expressed in terms of these two:
\begin{eqnarray}
\label{eq:identity as square of G_1 squared is G_3 fourth}
|1\rangle&=&|i\mathbf e_1\rangle^2=|\mathbf e_{0+3i}\rangle^4,\\
\label{eq:ie_3 is G_3 squared}
|i\mathbf e_3\rangle&=&|\mathbf e_{0+3i}\rangle^2=|i\mathbf e_3\rangle^{-1},\\
\label{eq:e_0-3i is G_3 cube is e_0+3i inverse}
|\mathbf e_{0-3i}\rangle&=&|\mathbf e_{0+3i}\rangle^3=|\mathbf e_{0+3i}\rangle^{-1},\\
\label{eq:ie_2 is G_1 G_3 squared is ie_2 inverse}
|i\mathbf e_2\rangle&=&|i\mathbf e_1\rangle|\mathbf e_{0+3i}\rangle^2=|i\mathbf e_2\rangle^{-1},\\
\label{eq:ie_1+2 is G_1 G_3 is ie_1+2 inverse}
|i\mathbf e_{1+2}\rangle&=&|i\mathbf e_1\rangle|\mathbf e_{0+3i}\rangle=|i\mathbf e_{1+2}\rangle^{-1},\\
\label{eq:ie_1-2 is G_1 G_3 cube is ie_1+2 inverse}
|i\mathbf e_{1-2}\rangle&=&|i\mathbf e_1\rangle|\mathbf e_{0+3i}\rangle^3=|i\mathbf e_{1-2}\rangle^{-1}.
\end{eqnarray}
(See Table 1)

\begin{table}
\footnotesize
\begin{center}
\begin{tabular}{|c||c|c|c|c|}
\hline
\ & $|1\rangle$\T & $|i\mathbf e_1\rangle$ & $|i\mathbf e_2\rangle$ & $|i\mathbf e_3\rangle$\\
\hline\hline
$\T|1\rangle$ & $|1\rangle$ & $|i\mathbf e_1\rangle$ & $|i\mathbf e_2\rangle$ & $|i\mathbf e_3\rangle$\\
\hline
$\T|i\mathbf e_1\rangle$ & $|i\mathbf e_1\rangle$ & $|1\rangle$ & $|i\mathbf e_3\rangle$ & $|i\mathbf e_2\rangle$\\
\hline
$\T|i\mathbf e_2\rangle$ & $|i\mathbf e_2\rangle$ & $|i\mathbf e_3\rangle$ & $|1\rangle$ & $|i\mathbf e_1\rangle$\\
\hline
$\T|i\mathbf e_3\rangle$ & $|i\mathbf e_3\rangle$ & $|i\mathbf e_2\rangle$ & $|i\mathbf e_1\rangle$ & $|1\rangle$\\
\hline
$\T|\mathbf e_{0+3i}\rangle$ & $|\mathbf e_{0+3i}\rangle$ & $|i\mathbf e_{1-2}\rangle$ & $|i\mathbf e_{1+2}\rangle$ & $|\mathbf e_{0-3i}\rangle$\\
\hline
$\T|\mathbf e_{0-3i}\rangle$ & $|\mathbf e_{0-3i}\rangle$ & $|i\mathbf e_{1+2}\rangle$ & $|i\mathbf e_{1-2}\rangle$ & $|\mathbf e_{0+3i}\rangle$\\
\hline
$\T|i\mathbf e_{1+2}\rangle$ & $|i\mathbf e_{1+2}\rangle$ & $|\mathbf e_{0-3i}\rangle$ & $|\mathbf e_{0+3i}\rangle$ & $|i\mathbf e_{1-2}\rangle$\\
\hline
$\T|i\mathbf e_{1-2}\rangle$ & $|i\mathbf e_{1-2}\rangle$ & $|\mathbf e_{0+3i}\rangle$ & $|\mathbf e_{0-3i}\rangle$ & $|i\mathbf e_{1+2}\rangle$\\
\hline
\end{tabular}
\bigskip

\begin{tabular}{|c||c|c|c|c|}
\hline
\ & $|\mathbf e_{0+3i}\rangle$\T & $|\mathbf e_{0-3i}\rangle$ & $|i\mathbf e_{1+2}\rangle$ & $|i\mathbf e_{1-2}\rangle$\\
\hline\hline
$\T|1\rangle$ & $|\mathbf e_{0+3i}\rangle$ & $|\mathbf e_{0-3i}\rangle$ & $|i\mathbf e_{1+2}\rangle$ & $|i\mathbf e_{1-2}\rangle$\\
\hline
$\T|i\mathbf e_1\rangle$ & $|i\mathbf e_{1+2}\rangle$ & $|i\mathbf e_{1-2}\rangle$ & $|\mathbf e_{0+3i}\rangle$ & $|\mathbf e_{0-3i}\rangle$\\
\hline
$\T|i\mathbf e_2\rangle$ & $|i\mathbf e_{1-2}\rangle$ & $|i\mathbf e_{1+2}\rangle$ & $|\mathbf e_{0-3i}\rangle$ & $|\mathbf e_{0+3i}\rangle$\\
\hline
$\T|i\mathbf e_3\rangle$ & $|\mathbf e_{0-3i}\rangle$ & $|\mathbf e_{0+3i}\rangle$ & $|i\mathbf e_{1-2}\rangle$ & $|i\mathbf e_{1+2}\rangle$\\
\hline
$\T|\mathbf e_{0+3i}\rangle$ & $|i\mathbf e_3\rangle$ & $|1\rangle$ & $|i\mathbf e_1\rangle$ & $|i\mathbf e_2\rangle$\\
\hline
$\T|\mathbf e_{0-3i}\rangle$ & $|1\rangle$ & $|i\mathbf e_3\rangle$ & $|i\mathbf e_2\rangle$ & $|i\mathbf e_1\rangle$\\
\hline
$\T|i\mathbf e_{1+2}\rangle$ & $|i\mathbf e_2\rangle$ & $|i\mathbf e_1\rangle$ & $|1\rangle$ & $|i\mathbf e_3\rangle$\\
\hline
$\T|i\mathbf e_{1-2}\rangle$ & $|i\mathbf e_1\rangle$ & $|i\mathbf e_2\rangle$ & $|i\mathbf e_3\rangle$ & $|1\rangle$\\
\hline
\end{tabular}

\end{center}
\caption{\footnotesize Multiplication tables for the dihedral group $\mathcal D_4$.}
\end{table}

\subsection{Identities}

In actual computations, it is simpler to work with the following five group elements: $|i\mathbf e_1\rangle$, $|i\mathbf e_2\rangle$, $|i\mathbf e_3\rangle$, $|i\mathbf e_{0+3i}\rangle$, and $|i\mathbf e_{0-3i}\rangle$.  The first three correspond to $180^\circ$ flips about a coordinate axis; the next two correspond to counterclockwise and clockwise quarter rotations about the $z-$axis direction $\mathbf e_3$.

\textbf{a.  Products.}  The flip operators $|i\mathbf e_1\rangle$, $|i\mathbf e_2\rangle$, $|i\mathbf e_3\rangle$ are generators of an Abelian (commuting) quasi-quaternion algebra:
\begin{equation}
\label{eq:quasiquaternion ie_1 ie_2 ie_3 squared}
|\,i\mathbf e_1\,\rangle^2=|\,i\mathbf e_2\,\rangle^2=|\,i\mathbf e_3\,\rangle^2=|\,1\,\rangle,
\end{equation}
and
\begin{eqnarray}
\label{eq:quasiquaternion ie_1 ie_2 is ie_3}
|\,i\mathbf e_1\,\rangle|\,i\mathbf e_2\,\rangle=|\,i\mathbf e_3\,\rangle=|\,i\mathbf e_2\,\rangle|\,i\mathbf e_1\,\rangle,\\
\label{eq:quasiquaternion ie_2 ie_3 is ie_1}
|\,i\mathbf e_2\,\rangle|\,i\mathbf e_3\,\rangle=|\,i\mathbf e_1\,\rangle=|\,i\mathbf e_3\,\rangle|\,i\mathbf e_2\,\rangle,\\
\label{eq:quasiquaternion ie_3 ie_1 is ie_2}
|\,i\mathbf e_3\,\rangle|\,i\mathbf e_1\,\rangle=|\,i\mathbf e_2\,\rangle=|\,i\mathbf e_1\,\rangle|\,i\mathbf e_3\,\rangle.
\end{eqnarray}
Equation~(\ref{eq:quasiquaternion ie_1 ie_2 ie_3 squared}) states that a $180^\circ-$rotation applied twice leaves an object unchanged; Eqs.~(\ref{eq:quasiquaternion ie_1 ie_2 is ie_3}) to (\ref{eq:quasiquaternion ie_3 ie_1 is ie_2}) state that two consecutive $180^\circ\!-$rotations about two mutually orthogonal axes is equal to a $180^\circ\!-$rotation about an axis perpendicular to the previous two.

The product of the flip operator $|i\mathbf e_k\rangle$ and the quarter rotation operator $|i\mathbf e_{0\pm 3i}\rangle$ satisfies the following conjugation-commutation relations:
\begin{eqnarray}
\label{eq:conjugation relation ie_1 and e_0+3i}
|\,i\mathbf e_1\,\rangle |\,\mathbf e_{0\pm 3i}\,\rangle
&=&\frac{1}{\sqrt{2}}\, |\,i\mathbf e_1(1\pm i\mathbf e_3)\,\rangle \nonumber\\
&=&\frac{1}{\sqrt{2}}\, |(1\mp i\mathbf e_3)i\mathbf e_1\,\rangle  \nonumber\\
&=&|\,\mathbf e_{0\mp 3i}\,\rangle |\,i\mathbf e_1\,\rangle,\\
\label{eq:conjugation relation ie_2 and e_0+3i}
|\,i\mathbf e_2\,\rangle |\,\mathbf e_{0\pm 3i}\,\rangle
&=&|\,\mathbf e_{0\mp 3i}\,\rangle |\,i\mathbf e_2\,\rangle,\\
\label{eq:conjugation relation ie_3 and e_0+3i}
|\,i\mathbf e_3\,\rangle |\,\mathbf e_{0\pm 3i}\,\rangle
&=&|\,\mathbf e_{0\pm 3i}\,\rangle |\,i\mathbf e_3\,\rangle.
\end{eqnarray}
The passing over of $|\,i\mathbf e_1\,\rangle$ or $|\,i\mathbf e_1\,\rangle$ transforms $|\,\mathbf e_{0+3i}\,\rangle$ into its conjugate $|\,\mathbf e_{0-3i}\,\rangle$.  But the passing over of $|\,i\mathbf e_3\,\rangle$ leaves $|\,\mathbf e_{0+3i}\,\rangle$ unchanged.  

Alternatively, we can impose that $|\mathbf e_{0\pm 3i}\rangle$ remain invariant and let $|i\mathbf e_j\rangle$ mutate, for $j=1,2$.  That is, we write
\begin{equation}
\label{eq:e_0 pm 3i ie_j is ie_j' e_0 pm 3i}
|\mathbf e_{0\pm 3i}\rangle|i\mathbf e_j\rangle=|i\mathbf e_j'\rangle|\mathbf e_{0\pm 3i}\rangle.
\end{equation}
Our problem is to express $|i\mathbf e_j'\rangle$ in terms of $|i\mathbf e_j\rangle$.

Let us first consider $|\mathbf e_{0\pm 3i}\rangle$.  From Eqs.~(\ref{eq:conjugation relation ie_1 and e_0+3i}) and (\ref{eq:conjugation relation ie_3 and e_0+3i}), we have
\begin{equation}
\label{eq:e_0+3i ie_j is ie_j e_0-3i}
|\,\mathbf e_{0+3i}\rangle|i\mathbf e_j\rangle=|i\mathbf e_j\rangle|\,\mathbf e_{0-3i}\rangle.
\end{equation}
Since
\begin{equation}
\label{eq:e_0-3i is ie_3 e_0+3i}
|\,\mathbf e_{0-3i}\rangle=|\,\mathbf e_{0+3i}\rangle^2|\,\mathbf e_{0+3i}\rangle=|i\mathbf e_3\rangle|\,\mathbf e_{0+3i}\rangle,
\end{equation}
then Eq.~(\ref{eq:e_0+3i ie_j is ie_j e_0-3i}) becomes
\begin{equation}
\label{eq:e_0+3i ie_j is ie_j ie_3 e_0+3i}
|\,\mathbf e_{0+3i}\rangle|i\mathbf e_j\rangle=|i\mathbf e_j\rangle|i\mathbf e_3\rangle|\,\mathbf e_{0+3i}\rangle,
\end{equation}
so that
\begin{equation}
\label{eq:ie_j' is ie_j ie_3}
|i\mathbf e_j'\rangle=|i\mathbf e_j\rangle|i\mathbf e_3\rangle.
\end{equation}

Similarly, for $|\,\mathbf e_{0-3i}\rangle$, we have
\begin{equation}
\label{eq:e_0-3i ie_k is ie_k e_0+3i}
|\,\mathbf e_{0-3i}\rangle|i\mathbf e_j\rangle=|i\mathbf e_j\rangle|\,\mathbf e_{0+3i}\rangle.
\end{equation}
Since 
\begin{eqnarray}
\label{eq:e_0+3i is ie_3 e_0-3i}
|\mathbf e_{0+3i}\rangle&=&|\mathbf e_{0+3i}\rangle|\mathbf e_{0-3i}\rangle^{-1}|\mathbf e_{0-3i}\rangle\\
&=&|\,\mathbf e_{0+3i}\rangle^2|\mathbf e_{0-3i}\rangle=|i\mathbf e_3\rangle|\mathbf e_{0-3i}\rangle,
\end{eqnarray}
then Eq.~(\ref{eq:e_0-3i ie_k is ie_k e_0+3i}) becomes
\begin{equation}
\label{eq:e_0-3i ie_j is ie_j ie_3 e_0-3i}
|\,\mathbf e_{0-3i}\rangle|i\mathbf e_j\rangle=|i\mathbf e_j\rangle|i\mathbf e_3\rangle|\,\mathbf e_{0-3i}\rangle,
\end{equation}
so that Eq.~(\ref{eq:ie_j' is ie_j ie_3}) still holds.

We may now combine Eqs.~(\ref{eq:e_0+3i ie_j is ie_j ie_3 e_0+3i}) and (\ref{eq:e_0-3i ie_j is ie_j ie_3 e_0-3i}) into one:
\begin{equation}
\label{eq:e_0pm3i ie_j is ie_j ie_3 e_0pm3i}
|\,\mathbf e_{0\pm 3i}\rangle|i\mathbf e_j\rangle=|i\mathbf e_j\rangle|i\mathbf e_3\rangle|\,\mathbf e_{0\pm 3i}\rangle.
\end{equation}
That is,
\begin{eqnarray}
\label{eq:e_0pm3i ie_1 is ie_2 e_0pm3i}
|\,\mathbf e_{0\pm 3i}\rangle|i\mathbf e_1\rangle&=&|i\mathbf e_2\rangle|\,\mathbf e_{0\pm 3i}\rangle,\\
\label{eq:e_0pm3i ie_2 is ie_1 e_0pm3i}
|\,\mathbf e_{0\pm 3i}\rangle|i\mathbf e_2\rangle&=&|i\mathbf e_1\rangle|\,\mathbf e_{0\pm 3i}\rangle.
\end{eqnarray}
Thus, the passing over of $|\,\mathbf e_{0\pm 3i}\rangle$ changes $|i\mathbf e_1\rangle$ to $|i\mathbf e_2\rangle$ and vice versa.

As a check, let us expand both sides of Eq.~(\ref{eq:e_0pm3i ie_1 is ie_2 e_0pm3i}):
\begin{eqnarray}
\label{eq:e_0pm3i ie_1 is ie_1 mp ie_2}
|\,\mathbf e_{0\pm 3i}\rangle|i\mathbf e_1\rangle&=&|\frac{1}{\sqrt{2}}(1\pm i\mathbf e_3)(i\mathbf e_1)\rangle,\nonumber\\
&=&|\frac{1}{\sqrt{2}}(i\mathbf e_1\mp i\mathbf e_2)\rangle,\\
|i\mathbf e_2\rangle|\,\mathbf e_{0\pm 3i}\rangle&=&|\frac{1}{\sqrt{2}}(i\mathbf e_2)(1\pm i\mathbf e_3)\rangle,\nonumber\\
&=&|\frac{1}{\sqrt{2}}(i\mathbf e_2\mp i\mathbf e_1)\rangle.
\end{eqnarray}
If we take the + case, then we have verified our claim.  If we choose the $-$ case, the arguments of the resulting expansions are additive inverses of each other, which means that the ket representations are equivalent by Hestenes's theorem in Eq.~(\ref{eq:rotation counterclockwise clockwise reduced}).  This ends the proof.

The proof for Eq.~(\ref{eq:e_0pm3i ie_2 is ie_1 e_0pm3i}) is similar.

\textbf{c.  Sums.}  Let $\mathbf r$ be the position of a point in three-dimensional Cartesian space:
\begin{equation}
\label{eq:r is x_1e_1 + x_2e_2}
\mathbf r=x_1\mathbf e_1+x_2\mathbf e_2+x_3\mathbf e_3.
\end{equation}
The actions of the $\mathcal D_4$ elements on $\mathbf r$ are 
\begin{eqnarray}
\label{eq:r 1}
\mathbf r\,|\,1\,\rangle&=&x_1\mathbf e_1+x_2\mathbf e_2+x_3\mathbf e_3,\\
\label{eq:r ie_1}
\mathbf r\,|\,i\mathbf e_1\,\rangle&=&x_1\mathbf e_1-x_2\mathbf e_2-x_3\mathbf e_3,\\
\label{eq:r ie_2}
\mathbf r\,|\,i\mathbf e_2\,\rangle&=&-x_1\mathbf e_1+x_2\mathbf e_2-x_3\mathbf e_3,\\
\label{eq:r ie_3}
\mathbf r\,|\,i\mathbf e_3\,\rangle&=&-x_1\mathbf e_1-x_2\mathbf e_2+x_3\mathbf e_3,\\
\label{eq:r e_0+3i}
\mathbf r\,|\,\mathbf e_{0+3i}\,\rangle&=&x_1\mathbf e_2-x_2\mathbf e_1+x_3\mathbf e_3,\\
\label{eq:r e_0-3i}
\mathbf r\,|\,\mathbf e_{0-3i}\,\rangle&=&-x_1\mathbf e_2+x_2\mathbf e_1+x_3\mathbf e_3,\\
\label{eq:r ie_1+2}
\mathbf r\,|\,i\mathbf e_{1+2}\,\rangle&=&x_1\mathbf e_2+x_2\mathbf e_1-x_3\mathbf e_3,\\
\label{eq:r ie_1-2}
\mathbf r\,|\,i\mathbf e_{1-2}\,\rangle&=&-x_1\mathbf e_2-x_2\mathbf e_1-x_3\mathbf e_3.
\end{eqnarray}

We may add the actions of the $\mathcal D_4$ elements to arrive at null results, depending on the definition of the position vector $\mathbf r$.  If we define
\begin{eqnarray}
\label{eq:r_1 is x_1e_1}
\mathbf r_1&=&x_1\mathbf e_1,\\
\label{eq:r_1+2 is x_1e_1 + x_2e_1}
\mathbf r_{1+2}&=&x_1\mathbf e_1+x_2\mathbf e_2,\\
\label{eq:r_1+2+3 is x_1e_1 + x_2e_2 + x_3e_3}
\mathbf r_{1+2+3}&=&x_1\mathbf e_1+x_2\mathbf e_2+x_3\mathbf e_3,
\end{eqnarray}
then we have the following quasi-quaternion identities:
\begin{eqnarray}
\label{eq:r_1 by 1 - ie_1 is 0}
0&=&\mathbf r_1(|1\rangle-|i\mathbf e_1\rangle),\\
\label{eq:r_1+2 by 1 + ie_3 is 0}
0&=&\mathbf r_{1+2}(|1\rangle+|i\mathbf e_3\rangle),\\
\label{eq:r_1+2 by ie_1 + ie_2 is 0}
0&=&\mathbf r_{1+2}(|i\mathbf e_1\rangle+|i\mathbf e_2\rangle),\\
\label{eq:r_1+2 by e_0+3i + e_0-3i is 0}
0&=&\mathbf r_{1+2}(|\mathbf e_{0+3i}\rangle+|\mathbf e_{0-3i}\rangle),\\
\label{eq:r_1+2 by ie_1+2 + ie_1-2 is 0}
0&=&\mathbf r_{1+2}(|i\mathbf e_{1+2}\rangle+|i\mathbf e_{1-2}\rangle),\\
\label{eq:r_1+2+3 by 1 + ie_1 + ie_2 + ie_3 is 0}
0&=&\mathbf r_{1+2+3}(|1\rangle+|i\mathbf e_1\rangle+|i\mathbf e_2\rangle+|i\mathbf e_3\rangle),
\end{eqnarray}
which we may rewrite as
\begin{eqnarray}
\label{eq:1 - ie_1 equiv 0 if r in Cl_1,0}
|1\rangle-|i\mathbf e_1\rangle&\equiv& 0;\quad \mathbf r\in \mathcal Cl_{1,0},\\ 
\label{eq:1 + ie_3 equiv 0 if r in Cl_2,0}
|1\rangle+|i\mathbf e_3\rangle&\equiv&0;\quad\mathbf r\in\mathcal Cl_{2,0},\\
\label{eq:ie_1 + ie_2 equiv 0 if r in Cl_2,0}
|i\mathbf e_1\rangle+|i\mathbf e_2\rangle&\equiv&0;\quad\mathbf r\in\mathcal Cl_{2,0},\\
\label{eq:e_0+3i + e_0-3i equiv 0 if r in Cl_2,0}
|\mathbf e_{0+3i}\rangle+|\mathbf e_{0-3i}\rangle&\equiv&0;\quad\mathbf r\in\mathcal Cl_{2,0},\\
\label{eq:ie_1+2 + ie_1-2 equiv 0 if r in Cl_2,0}
|i\mathbf e_{1+2}\rangle+|i\mathbf e_{1-2}\rangle&\equiv&0;\quad\mathbf r\in\mathcal Cl_{2,0},\\
\label{eq:1 + ie_1 + ie_2 + ie_3 equiv 0 if r in Cl_3,0}
|1\rangle+|i\mathbf e_1\rangle+|i\mathbf e_2\rangle+|i\mathbf e_3\rangle&\equiv&0;\quad\mathbf r\in\mathcal Cl_{3,0}.
\end{eqnarray}
Note that the ket operator identities that hold for a higher dimensional definition of the position vector $\mathbf r$ also hold for the lower dimensional definition.

\textbf{c. Products of Sums.}  Using the quasiquaternion algebra for orthogonal flips in Eqs.~(\ref{eq:quasiquaternion ie_1 ie_2 ie_3 squared}) to (\ref{eq:quasiquaternion ie_3 ie_1 is ie_2}), we can show that
\begin{equation}
\label{eq:1 +ie_1 squared is 1 + ie_1}
(|1\rangle+|i\mathbf e_k\rangle)^2=2(|1\rangle+|i\mathbf e_k\rangle),
\end{equation}
for $k=1,2,3$.  We can also show that   
\begin{eqnarray}
\label{eq:1 + ie_1 times 1 + ie_2 is 0}
0&=&(|1\rangle+|i\mathbf e_1\rangle)(|1\rangle+|i\mathbf e_2\rangle),\\
\label{eq:1 + ie_1 times 1 + ie_3 is 0}
&=&(|1\rangle+|i\mathbf e_1\rangle)(|1\rangle+|i\mathbf e_3\rangle),\\
\label{eq:1 + ie_2 times 1 + ie_3 is 0}
&=&(|1\rangle+|i\mathbf e_2\rangle)(|1\rangle+|i\mathbf e_3\rangle).
\end{eqnarray}
where we used the null identity in Eq.~(\ref{eq:1 + ie_1 + ie_2 + ie_3 equiv 0 if r in Cl_3,0}).  Note that these products are commutative. 

Another set of products that will become useful in the next subsection on dyadic operators are the following:
\begin{eqnarray}
\label{eq:e_0pm3i times 1 + ie_1 is 1 + ie_2 e_0pm3i}
|\,\mathbf e_{0\pm 3i}\rangle(|1\rangle+|i\mathbf e_1\rangle)=(|1\rangle+|i\mathbf e_2\rangle)|\,\mathbf e_{0\pm 3i}\rangle,\\
\label{eq:e_0pm3i times 1 + ie_2 is 1 + ie_1 e_0pm3i}
|\,\mathbf e_{0\pm 3i}\rangle(|1\rangle+|i\mathbf e_2\rangle)=(|1\rangle+|i\mathbf e_1\rangle)|\,\mathbf e_{0\pm 3i}\rangle,\\
\label{eq:e_0pm3i times 1 + ie_3 is 1 + ie_3 e_0pm3i}
|\,\mathbf e_{0\pm 3i}\rangle(|1\rangle+|i\mathbf e_3\rangle)=(|1\rangle+|i\mathbf e_3\rangle)|\,\mathbf e_{0\pm 3i}\rangle,
\end{eqnarray}
where we used the identities in Eqs.~(\ref{eq:conjugation relation ie_3 and e_0+3i}), (\ref{eq:e_0pm3i ie_1 is ie_2 e_0pm3i}), and (\ref{eq:e_0pm3i ie_2 is ie_1 e_0pm3i}).  Notice that only the binomial involving $|i\mathbf e_3\rangle$ is unaffected.

\subsection{Dyadics}

\textbf{a.  Right-Acting Operators.}  We may add the actions of the $\mathcal D_4$ elements to arrive at new transformations, which we shall denote using the $\cdot\,\textsf e_{\mu\nu}$ operators:
\begin{eqnarray}
\label{eq:r e_11 is x_1 e_1}
\mathbf r\cdot\textsf{e}_{11}&=&\frac{1}{2}\mathbf r(|1\rangle+|i\mathbf e_1\rangle)=x_1\mathbf e_1,\\
\label{eq:r e_12 is x_1 e_2}
\mathbf r\cdot\textsf{e}_{12}&=&\frac{1}{2}\mathbf r(|\,\mathbf e_{0+3i}\,\rangle+|i\mathbf e_{1+2}\rangle)\nonumber\\
&=&\frac{1}{2}\mathbf r(|1\rangle+|i\mathbf e_1\rangle)|\,\mathbf e_{0+3i}\rangle=x_1\mathbf e_2,\\
\label{eq:r e_21 is x_2 e_1}
\mathbf r\cdot\textsf{e}_{21}&=&\frac{1}{2}\mathbf r(|\,\mathbf e_{0-3i}\rangle+|i\mathbf e_{1+2}\rangle)\nonumber\\
&=&\frac{1}{2}\mathbf r(|1\rangle+|\,i\mathbf e_2\,\rangle)|\,\mathbf e_{0-3i}\rangle=x_2\mathbf e_1,\\
\label{eq:r e_22 is x_2 e_2}
\mathbf r\cdot\textsf{e}_{22}&=&\frac{1}{2}\mathbf r(|1\rangle+|i\mathbf e_2\rangle)=x_2\mathbf e_2.
\end{eqnarray}
Thus, the right-acting operator $\cdot\,\textsf e_{\mu\nu}$ extracts the $\mu-$component of $\mathbf r$ and replaces $\mathbf e_\mu$ by $\mathbf e_\nu$.

Formally, let us define the $\cdot\,\textsf e_{\mu\nu}$ operators in terms of the flip and quarter rotation operators:
\begin{eqnarray}
\label{eq:e_11 is half 1 + ie_1 right}
\cdot\,\textsf e_{11}&=&\frac{1}{2}(|1\rangle+|i\mathbf e_1\rangle),\\
\label{eq:e_12 is half 1 + ie_1 times e_0+3i right}
\cdot\,\textsf e_{12}&=&\frac{1}{2}(|1\rangle+|i\mathbf e_1\rangle)|\,\mathbf e_{0+3i}\rangle,\\
\label{eq:e_21 is half 1 + ie_2 times e_0-3i right}
\cdot\,\textsf e_{21}&=&\frac{1}{2}(|1\rangle+|i\mathbf e_2\rangle)|\,\mathbf e_{0-3i}\rangle,\\
\label{eq:e_22 is half 1 + ie_2 right}
\cdot\,\textsf e_{22}&=&\frac{1}{2}(|1\rangle+|i\mathbf e_2\rangle).
\end{eqnarray}
We shall refer to the $\cdot\,\textsf e_{\mu\nu}$ operators as the unit extraction-replacement operators.

The action of $\cdot\,\textsf e_{\mu\nu}$ on a vector $\mathbf e_\lambda$  and on another operator $\cdot\,\textsf e_{\mu'\nu'}$ are given by
\begin{eqnarray}
\label{eq:e_lambda e_mu nu right}
\mathbf e_{\lambda}\cdot\textsf e_{\mu\nu}&=&\delta_{\lambda\mu}\mathbf e_\nu,\\
\label{eq:e_mu' nu' e_mu nu right}
\cdot\,\textsf e_{\mu'\nu'}\cdot\textsf e_{\mu\nu}&=&\delta_{\nu'\mu}\cdot\textsf e_{\mu'\nu}.
\end{eqnarray}
The proof of Eq.~(\ref{eq:e_lambda e_mu nu right}) is obvious from Eqs.~(\ref{eq:r e_11 is x_1 e_1}) to (\ref{eq:r e_22 is x_2 e_2}).  For products involving $\cdot\,\textsf e_{11}$ and $\cdot\,\textsf e_{22}$ with other $\cdot\,\textsf e_{\mu\nu}$ operators, the identities in Eqs.~(\ref{eq:1 +ie_1 squared is 1 + ie_1}) and (\ref{eq:1 + ie_2 times 1 + ie_3 is 0}) are sufficient for the proof.  But for products involving $\cdot\,\textsf e_{12}$ and $\cdot\,\textsf e_{21}$, we need the additional identities in Eq.~(\ref{eq:e_0pm3i times 1 + ie_1 is 1 + ie_2 e_0pm3i}) and (\ref{eq:e_0pm3i times 1 + ie_2 is 1 + ie_1 e_0pm3i}), together with the theorem $|\mathbf e_{0-3i}\rangle=|\mathbf e_{0+3i}\rangle^{-1}$ in Eq.~(\ref{eq:e_0-3i is G_3 cube is e_0+3i inverse}).  

Notice the similarity of the action of the $\cdot\,\textsf e_{\mu\nu}$ operators to Symon's unit dyadic operators\cite{Symon_1971_Mechanics_p406}.  In fact, we could make the formal replacement
\begin{equation}
\label{eq:e_munu is e_mu e_nu right}
\cdot\,\textsf e_{\mu\nu}=\,\cdot\,\mathbf e_\mu\mathbf e_\nu
\end{equation}
and rederive the results in Eq.~(\ref{eq:e_lambda e_mu nu right}) and (\ref{eq:e_mu' nu' e_mu nu right}), assuming that the dot product takes precedence over the geometric product.

\textbf{b.  Left-Acting Operators.}  
The bra-to-ket conversion in Eq.~(\ref{eq:r prime is exp -itheta2 r exp itheta2 bra ket}), together with the equivalent ket representations in Eq.~(\ref{eq:D_4 identity 1}) to (\ref{eq:D_4 e_1 - e_2}), lets us write
\begin{eqnarray}
\label{eq:r 1 to left}
\mathbf r\,|\,1\,\rangle=&\langle\,(1)^{-1}\,|\,\mathbf r&=\langle \,1\,|\,\mathbf r,\\
\label{eq:r ie_1 to left}
\mathbf r\,|\,i\mathbf e_1\,\rangle=&\langle\,(i\mathbf e_1)^{-1}\,|\,\mathbf r&=\langle\,i\mathbf e_1\,|\,\mathbf r,\\
\label{eq:r ie_2 to left}
\mathbf r\,|\,i\mathbf e_2\,\rangle=&\langle\,(i\mathbf e_2)^{-1}\,|\,\mathbf r&=\langle\,i\mathbf e_2\,|\,\mathbf r,\\
\label{eq:r ie_3 to left}
\mathbf r\,|\,i\mathbf e_3\,\rangle=&\langle\,(i\mathbf e_3)^{-1}\,|\,\mathbf r&=\langle\,i\mathbf e_3\,|\,\mathbf r,\\
\label{eq:r e_0+3i to left}
\mathbf r\,|\,\mathbf e_{0+3i}\,\rangle=&\langle\,(e^{i\mathbf e_3\pi/4})^{-1}\,|\,\mathbf r&=\langle\,\mathbf e_{0-3i}\,|\,\mathbf r,\\
\label{eq:r e_0-3i to left}
\mathbf r\,|\,\mathbf e_{0-3i}\,\rangle=&\langle\,(e^{-i\mathbf e_3\pi/4})^{-1}\,|\,\mathbf r&=\langle\,\mathbf e_{0+3i}\,|\,\mathbf r,\\
\label{eq:r ie_1+2 to left}
\mathbf r\,|\,i\mathbf e_{1+2}\,\rangle=&\langle\,(e^{i(\mathbf e_{1+2})\pi/2})^{-1}\,|\,\mathbf r&=\langle\,i\mathbf e_{1+2}\,|\,\mathbf r,\\
\label{eq:r ie_1-2 to left}
\mathbf r\,|\,i\mathbf e_{1-2}\,\rangle=&\langle\,(e^{i(\mathbf e_{1-2})\pi/2})^{-1}\,|\,\mathbf r&=\langle\,i\mathbf e_{1-2}\,|\,\mathbf r.
\end{eqnarray}
The right side of Eqs.~(\ref{eq:r 1 to left}) to (\ref{eq:r ie_1-2 to left}) are the bra representations of the $\mathcal D_4$ group elements.

Using these ket-bra translations, we may transform Eqs.~(\ref{eq:r e_11 is x_1 e_1}) to (\ref{eq:r e_22 is x_2 e_2}) into
\begin{eqnarray}
\label{eq:e_11 r is x_1 e_1}
\textsf e_{11}\cdot\mathbf r&=&x_1\mathbf e_1,\\
\label{eq:e_21 r is x_1 e_1}
\textsf e_{21}\cdot\mathbf r&=&x_1\mathbf e_2,\\
\label{eq:e_12 r is x_2 e_1}
\textsf e_{12}\cdot\mathbf r&=&x_2\mathbf e_1,\\
\label{eq:e_22 r is x_2 e_2}
\textsf e_{22}\cdot\mathbf r&=&x_2\mathbf e_2,
\end{eqnarray}
where
\begin{eqnarray}
\label{eq:e_11 is 1 + ie_1 left}
\textsf e_{11}\cdot&=&\frac{1}{2}(\langle 1|+\langle i\mathbf e_1|),\\
\label{eq:e_21 is e_0-3i times 1 + ie_1 left}
\textsf e_{21}\cdot&=&\frac{1}{2}\langle\mathbf e_{0-3i}|(\langle 1|+\langle i\mathbf e_1|),\\
\label{eq:e_12 is e_0+3i times 1 + ie_2 left}
\textsf e_{12}\cdot&=&\frac{1}{2}\langle\mathbf e_{0+3i}|(\langle 1|+\langle i\mathbf e_2|),\\
\label{eq:e_22 is 1 + ie_2 left}
\textsf e_{22}\cdot&=&\frac{1}{2}(\langle 1|+\langle i\mathbf e_2|)
\end{eqnarray}
are the four left-acting dyadic operators.

Using similar theorems as those used to prove Eqs.~(\ref{eq:e_lambda e_mu nu right}) and (\ref{eq:e_mu' nu' e_mu nu right}), we can show that
\begin{eqnarray}
\label{eq:e_numu e_lambda left}
\textsf e_{\nu\mu}\cdot\mathbf e_\lambda&=&\delta_{\mu\lambda}\mathbf e_\nu,\\
\label{eq:e_numu e_nu'mu' left}
\textsf e_{\nu\mu}\cdot\textsf e_{\nu'\mu'}\cdot&=&\delta_{\mu\nu'}\textsf e_{\nu\mu'}\cdot
\end{eqnarray}
Notice that the left- and right-acting diadics are related by transposition:
\begin{equation}
\label{eq:e_munu right transpose is e_numu left}
(\cdot\,\textsf e_{\mu\nu})^T=\textsf e_{\nu\mu}\cdot
\end{equation}

\subsection{Matrices}

\textbf{a.  Right-Acting Matrices.}  Let $\mathbf r$ and $\mathbf r^\prime$ be vectors on the $xy-$plane,
\begin{eqnarray}
\label{eq:r in Cl_2,0}
\mathbf r&=&x_1\mathbf e_1+x_2\mathbf e_2\in\mathcal Cl_{2,0},\\
\label{eq:r' in Cl_2,0}
\mathbf r^\prime&=&x_1^\prime\mathbf e_1+x_2^\prime\mathbf e_2\in\mathcal Cl_{2,0},
\end{eqnarray}
and let $\cdot\,\textsf M$ be a linear combination of right-acting dyadics,
\begin{equation}
\label{eq:M is M_11 e_11 + M_12 e_12 + M_21 e_21 + M_22 e_22 right}
\cdot\,\textsf M=\,\cdot\,\textsf e_{11}M_{11}+\,\cdot\,\textsf e_{12}M_{12}+\,\cdot\,\textsf e_{21}M_{21}+\,\cdot\,\textsf e_{22}M_{22},
\end{equation}
where
\begin{eqnarray}
\label{eq:e_11 and e_12 matrix right}
\cdot\,\textsf e_{11}=
\begin{pmatrix}
1&0\\
0&0
\end{pmatrix}
,&\quad&
\cdot\,\textsf e_{12}=
\begin{pmatrix}
0&1\\
0&0
\end{pmatrix}
,
\\
\label{eq:e_21 and e_22 matrix right}
\cdot\,\textsf e_{21}=
\begin{pmatrix}
0&0\\
1&0
\end{pmatrix}
,&\quad&
\cdot\,\textsf e_{22}=
\begin{pmatrix}
0&0\\
0&1
\end{pmatrix}
,
\end{eqnarray}
are the four Fermion matrices.

If $\mathbf r^\prime$ is vector $\mathbf r$ right-acted by $\cdot\,\textsf M$, 
\begin{equation}
\label{eq:r' is r M}
\mathbf r^\prime=\mathbf r\cdot\textsf M,
\end{equation}
then by Eq.~(\ref{eq:e_lambda e_mu nu right}), we have 
\begin{eqnarray}
\label{eq:r' is r M e_1 part}
x_1^\prime=x_1M_{11}+x_2M_{21},\\
\label{eq:r' is r M e_2 part}
x_2^\prime=x_1M_{12}+x_2M_{22}.
\end{eqnarray}
In matrix form, this is
\begin{equation}
\label{eq:r' is r M matrix}
\begin{pmatrix}
x_1'\\
x_2'
\end{pmatrix}
=
\begin{pmatrix}
x_1\\
x_2
\end{pmatrix}
\begin{pmatrix}
M_{11}&M_{12}\\
M_{21}&M_{22}
\end{pmatrix}
.
\end{equation}
Notice that to get $x_1'$ we multiply the $\mathbf r$ column with the left column of $\cdot\,\textsf M$, as prescribed by Eq.~(\ref{eq:r' is r M e_1 part}); for $x_2'$, we multiply $\mathbf r$ with the right column of $\cdot\,\textsf M$, as prescribed by Eq.~(\ref{eq:r' is r M e_2 part}).  

On the other hand, to derive the rules for the matrix products, let $\cdot\,\textsf M$, $\cdot\,\textsf M'$, and $\cdot\,\textsf M''$ be $2\times 2$  right-acting matrices such that
\begin{equation}
\label{eq:M'' is M M'}
\cdot\,\textsf M''=\,\cdot\textsf M\cdot\textsf M'.
\end{equation}
where $\textsf M$ is defined in Eq.~(\ref{eq:M is M_11 e_11 + M_12 e_12 + M_21 e_21 + M_22 e_22 right}); $\textsf M'$ and $\textsf M''$ are defined similarly.  That is,
\begin{equation}
\label{eq:M'' is M M' matrix form}
\begin{pmatrix}
M_{11}''&M_{12}''\\
M_{21}''&M_{22}''\\
\end{pmatrix}
=
\begin{pmatrix}
M_{11}&M_{12}\\
M_{21}&M_{22}\\
\end{pmatrix}
\begin{pmatrix}
M_{11}'&M_{12}'\\
M_{21}'&M_{22}'\\
\end{pmatrix}
\end{equation}
Using the rule for dyadic products in Eq.~(\ref{eq:e_mu' nu' e_mu nu right}) and separating the matrix components, we obtain
\begin{eqnarray}
\label{eq:M'' is M M' e_11 part}
M_{11}''&=&M_{11}M_{11}'+M_{12}M_{21}',\\
\label{eq:M'' is M M' e_12 part}
M_{12}''&=&M_{11}M_{12}'+M_{12}M_{22}',\\
\label{eq:M'' is M M' e_21 part}
M_{21}''&=&M_{21}M_{11}'+M_{22}M_{21}',\\
\label{eq:M'' is M M' e_22 part}
M_{22}''&=&M_{21}M_{12}'+M_{22}M_{22}',
\end{eqnarray}
which agrees with the traditional definition of matrix multiplication.

\textbf{b. Left-Acting Matrices.}  If $\textsf M$ is the right acting matrix defined in Eq.~(\ref{eq:M is M_11 e_11 + M_12 e_12 + M_21 e_21 + M_22 e_22 right}), then by Eqs.~(\ref{eq:e_munu right transpose is e_numu left}), we may convert $\textsf M$ into left-acting by taking its transpose, as noted in Symon\cite{Symon_1971_Mechanics_p408}:
\begin{equation}
\label{eq:r' is M transpose r is r M}
\mathbf r^\prime=\textsf M^T\cdot\mathbf r=\mathbf r\cdot\textsf M,
\end{equation}
where $\mathbf r$ and $\mathbf r^\prime$ are the two-dimensional vectors defined in Eq.~(\ref{eq:r in Cl_2,0}) to (\ref{eq:r' in Cl_2,0}), and
\begin{equation}
\label{eq:M transpose component form}
\textsf M^T=M_{11}\textsf e_{11}+M_{12}\textsf e_{21}+M_{21}\textsf e_{12}+M_{22}\textsf e_{22}.
\end{equation}
That is,
\begin{eqnarray}
\label{eq:x' is M transpose x is x M matrix form}
\begin{pmatrix}
x_1^\prime\\
x_2^\prime
\end{pmatrix}
&=&
\begin{pmatrix}
M_{11}&M_{21}\\
M_{12}&M_{22}
\end{pmatrix}
\begin{pmatrix}
x_1\\
x_2
\end{pmatrix}
\\
&=&
\begin{pmatrix}
x_1\\
x_2
\end{pmatrix}
\begin{pmatrix}
M_{11}&M_{12}\\
M_{21}&M_{22}
\end{pmatrix}
.
\end{eqnarray}
As expected, the first component $x_1'$ is the product of the first row of the left acting matrix $\textsf M^T$and the column vector $\mathbf x$; for the second component $x_2'$, the product is with the second row.  Notice that the column-column multiplication for the action of right-acting matrices on column vectors is pedagogically simpler than the row-column multiplication for left-acting matrices.

Applying another right-acting operator $\textsf M'$ on Eq.~(\ref{eq:r' is M transpose r is r M}) yields
\begin{equation}
\label{eq:M' transpose M transpose r is r M M'}
\textsf M'^T\cdot\textsf M^T\cdot\mathbf r=\mathbf r\cdot\textsf M\cdot\textsf M'.
\end{equation}
Because $\textsf M\textsf M'$ is a matrix, then by Eq.~(\ref{eq:r' is M transpose r is r M}), we have
\begin{equation}
(\cdot\,\textsf M\cdot\textsf M')^T=\textsf M'^T\cdot\textsf M^T\cdot,
\end{equation}
which is the known theorem for the transpose of a product of matrices\cite{Heading_1966_MatrixTheoryforPhysicists_p13}.  But unlike the traditional expansion, we clarify the handedness of the matrices.

The product of two left-acting matrices is the same as if the matrices were right acting, by virtue of the dyadic product relation in Eq.~(\ref{eq:e_numu e_nu'mu' left}).  That is, if we rewrite Eq.~(\ref{eq:M'' is M M'}) as
\begin{equation}
\label{eq:M'' is M M' left}
\textsf M''\cdot\,=\textsf M\cdot\textsf M'\cdot, 
\end{equation}
then we will still arrive at Eqs. (\ref{eq:M'' is M M' matrix form}) to (\ref{eq:M'' is M M' e_22 part}).

\textbf{c.  $\mathcal D_4$ Matrices.}  The $2\times 2$ matrix representations of the dihedral group $\mathcal D_4$ elements may be readily obtained by setting $x_3=0$ in Eqs.~(\ref{eq:r 1}) to (\ref{eq:r ie_1-2}), and using the rules for the actions of the Fermion dyadics on vectors:
\begin{eqnarray}
\label{eq:r 1 matrix}
\mathbf r|1\rangle&=&\mathbf r
\begin{pmatrix}
1&0\\
0&1
\end{pmatrix}
=\mathbf r\cdot(\textsf e_{11}+\textsf e_{22}),\\
\label{eq:r ie_1 matrix}
\mathbf r|i\mathbf e_1\rangle&=&\mathbf r
\begin{pmatrix}
1&0\\
0&-1
\end{pmatrix}
=\mathbf r\cdot(\textsf e_{11}-\textsf e_{22}),\\
\label{eq:r ie_2 matrix}
\mathbf r|i\mathbf e_2\rangle&=&\mathbf r
\begin{pmatrix}
-1&0\\
0&1
\end{pmatrix}
=\mathbf r\cdot(-\textsf e_{11}+\textsf e_{22}),\\
\label{eq:r ie_3 matrix}
\mathbf r|i\mathbf e_3\rangle&=&\mathbf r
\begin{pmatrix}
-1&0\\
0&-1
\end{pmatrix}
=\mathbf r\cdot(-\textsf e_{11}-\textsf e_{22}),\\
\label{eq:r e_0+3i matrix}
\mathbf r|\mathbf e_{0+3i}\rangle&=&\mathbf r
\begin{pmatrix}
0&1\\
-1&0
\end{pmatrix}
=\mathbf r\cdot(\textsf e_{12}-\textsf e_{21}),\\
\label{eq:r e_0-3i matrix}
\mathbf r|\mathbf e_{0-3i}\rangle&=&\mathbf r
\begin{pmatrix}
0&-1\\
1&0
\end{pmatrix}
=\mathbf r\cdot(-\textsf e_{12}+\textsf e_{21}),\\
\label{eq:r ie_1+2 matrix}
\mathbf r|i\mathbf e_{1+2}\rangle&=&\mathbf r
\begin{pmatrix}
0&1\\
1&0
\end{pmatrix}
=\mathbf r\cdot(\textsf e_{12}+\textsf e_{21}),\\
\label{eq:r ie_1-2 matrix}
\mathbf r|i\mathbf e_{1-2}\rangle&=&\mathbf r
\begin{pmatrix}
0&-1\\
-1&0
\end{pmatrix}
=\mathbf r\cdot(-\textsf e_{12}-\textsf e_{21}).
\end{eqnarray}
Note that the bra representations of these operators have the same arguments as those of the kets, except for the ket $|\mathbf e_{0\pm3i}\rangle$ whose bra representation is $\langle\mathbf e_{0\mp3i}|$, as noted in Eqs.~(\ref{eq:r 1 to left}) to (\ref{eq:r ie_1-2 to left}).   Note also that the transpose of these right-acting matrices are the left-acting matrix representations of the dihedral group $\mathcal D_4$ as given by Lakshminarayanan and Viana\cite{LakshminarayananViana_2005_josaa22i11pp2483-2489_p2484}.

\textbf{d. Pauli Matrices.}  In terms of leftt-acting bra operators, Campbell's four primary matrices $|\mathbf I|$, $|\!+\!|$, $|\mathbf x|$, and $|\mathbf J|$ and their negatives may be expressed as\cite{Campbell_1997_OptomVisSci74i6pp381-387_p383}
\begin{eqnarray}
\label{eq:1 is I}
\langle 1|&=&|\mathbf I|=
\begin{pmatrix}
1&0\\
0&1
\end{pmatrix}
\equiv-\langle i\mathbf e_3|,\\
\label{eq:ie_1 is +}
\langle i\mathbf e_1|&=&|\!+\!|=
\begin{pmatrix}
1&0\\
0&-1
\end{pmatrix}
\equiv-\langle i\mathbf e_2|,\\
\label{eq:ie_1+2 is x}
\langle i\mathbf e_{1+2}|&=&|\mathbf x|=
\begin{pmatrix}
0&1\\
1&0
\end{pmatrix}
\equiv-\langle i\mathbf e_{1-2}|,\\
\label{eq:ie_0-3i is J}
\langle \mathbf e_{0-3i}|&=&|\mathbf J|=
\begin{pmatrix}
0&1\\
-1&0
\end{pmatrix}
\equiv-\langle i\mathbf e_{0+3i}|.
\end{eqnarray}
Note that these equivalence relations are only valid if their operand $\mathbf r\in\mathcal Cl_{2,0}$, as required by the ket identities in Eqs.~(\ref{eq:1 + ie_3 equiv 0 if r in Cl_2,0}) to (\ref{eq:ie_1+2 + ie_1-2 equiv 0 if r in Cl_2,0}).

The Pauli $\hat\sigma-$matrices may be defined similarly:
\begin{eqnarray}
\label{eq:sigma_0 is 1 matrix}
\hat\sigma_0&=&
\begin{pmatrix}
1&0\\
0&1
\end{pmatrix}
=(\textsf e_{11}+\textsf e_{22})\,\cdot\,
\equiv\langle 1|,\\
\label{eq:sigma_1 is ie_1+2 matrix}
\hat\sigma_1&=&
\begin{pmatrix}
0&1\\
1&0
\end{pmatrix}
=(\textsf e_{12}+\textsf e_{21})\,\cdot\,\equiv\langle i\mathbf e_{1+2}|,\\
\label{eq:sigma_2 is ie_0-3i matrix}
\hat\sigma_2&=&
\begin{pmatrix}
0&-i\\
i&0
\end{pmatrix}
=(-i\textsf e_{12}+i\textsf e_{21})\,\cdot\,
\equiv i\langle \mathbf e_{0-3i}|,\\
\label{eq:sigma_3 is ie_1 matrix}
\hat\sigma_3&=&
\begin{pmatrix}
1&0\\
0&-1
\end{pmatrix}
=(\textsf e_{11}-\textsf e_{22})\,\cdot\,
\equiv \langle i\mathbf e_1|,
\end{eqnarray}
Note the extra imaginary number $i=\mathbf e_1\mathbf e_2\mathbf e_3$ in $\hat\sigma_2$.

To geometrically interpret the Pauli sigma matrices, we express it in terms of ket exponentials:
\begin{eqnarray}
\label{eq:sigma_0 r is r 1}
\hat\sigma_0\cdot\mathbf r&=&\mathbf r|1\rangle,\\
\label{eq:sigma_1 r is r ie_1+2}
\hat\sigma_1\cdot\mathbf r&=&\mathbf r|i\mathbf e_{1+2}\rangle=\mathbf r|e^{i\mathbf e_{1+2}\pi/2}\rangle,\\
\label{eq:sigma_2 r is r ie_0-3i}
\hat\sigma_2\cdot\mathbf r&=&(i\mathbf r)|\mathbf e_{0-3i}\rangle=(i\mathbf r)|e^{-i\mathbf e_3\pi/4}\rangle,\\
\label{eq:sigma_3 r is r ie_1}
\hat\sigma_3\cdot\mathbf r&=&\mathbf r|i\mathbf e_1\rangle=\mathbf r|e^{i\mathbf e_1\pi/2}\rangle.
\end{eqnarray}
Because the arguments of the exponentials are half-angles, then $\hat\sigma_0$ is an identity rotation of the vector 
\begin{equation}
\label{eq:r is x_1e_1 + x_2e_2 is column}
\mathbf r=x_1\mathbf e_1+x_2\mathbf e_2=
\begin{pmatrix}
x_1\\
x_2
\end{pmatrix}
,
\end{equation}
$\hat\sigma_1$ is a flip (rotation by $\pi$) about the the diagonal axis $\mathbf e_{1+2}$ and $\hat\sigma_3$ is a flip about the $\mathbf e_1$ axis; the matrix $\sigma_2$ is a $\pi/2$ rotation about $\mathbf e_3$ and the result multiplied by $i$. 

\section{Conclusions}

We represented vector rotation operators in terms of bras or kets of half-angle exponentials in Clifford (geometric) algebra $\mathcal Cl_{3,0}$.  The bras are left-acting operators with negative exponential arguments; the kets are right-acting operators with postive arguments.  We showed that $SO_3$ is a rotation group and the dihedral group $\mathcal D_4$ as its finite subgroup.  We derived several theorems: Euler-Rodrigues formulas, Hestenes's representation equivalence, bra-to-ket transformation, and ket commutation-conjugation identities. We computed the group table of $\mathcal D_4$ by illustrating a square on the $xy-$plane, labeling its several symmetry axis, and defining its half-angle exponential ket operators.  We took linear combinations of the elements of $\mathcal D_4$ to represent the four Fermion matrices, which in turn we used to decompose any $2\times 2$ matrix.  We showed that bra and ket operators generate left- and right-acting matrices, respectively, which are related by transposition.  We also showed that the Pauli spin matrices are not vectors but vector rotation operators, except for $\hat\sigma_2$ which required a subsequent multiplication by the imaginary number $i$ geometrically interpreted as the unit oriented volume.

The bra and ket representations may be used to describe the elements of other crystallographic point groups: the half-angle exponential representation would provide us immediate information on the direction of the symmetry axis and its corresponding rotation angles.  So in the computation of group tables, we need not worry about matrix representations of group elements: as long as we can imagine the products of composite rotations using palm-twisting arguments or compute them using the rules of geometric algebra, we would be able to arrive at our answers in the least time, even mentally.

We may also extend our geometric algebra reformulation to $3\times 3$ matrices.  In this case, the symmetry group that we must use is that of a cube.

\section*{\small{Acknowledgments}}
This research was supported by the Manila Observatory and by the Physics Department of Ateneo de Manila University.

\end{document}